# Requirements-Aided Automatic Test Case Generation for Industrial Cyber-physical Systems

Roopak Sinha[1], Cheng Pang[2], Gerardo Santillán Martínez[2], Juha Kuronen[3], and Valeriy Vyatkin[2,4]
[1]Computer and Mathematical Sciences, Auckland University of Technology, Auckland, New Zealand
[2]Department of Electrical Engineering and Automation, Aalto University, Helsinki, Finland
[3]Fortum Oyj, Helsinki, Finland
[4]Department of Computer Science, Electrical and Space Engineering, Luleå University of Technology, Sweden
emails: roopak.sinha@aut.ac.nz, cheng.pang.phd@ieee.org, gerardo.santillan@aalto.fi, juha.kuronen@fortum.com, vyatkin@ieee.org

**Abstract**

*Industrial cyber-physical systems require complex distributed software to orchestrate many heterogeneous mechatronic components and control multiple physical processes. Industrial automation software is typically developed in a model-driven fashion where abstractions of physical processes called plant models are co-developed and iteratively refined along with the control code. Testing such multi-dimensional systems is extremely difficult because often models might not be accurate, do not correspond accurately with subsequent refinements, and the software must eventually be tested on the real plant, especially in safety-critical systems like nuclear plants. This paper proposes a framework wherein high-level functional requirements are used to automatically generate test cases for designs at all abstraction levels in the model-driven engineering process. Requirements are initially specified in natural language and then analyzed and specified using a formalized ontology. The requirements ontology is then refined along with controller and plant models during design and development stages such that test cases can be generated automatically at any stage. A representative industrial water process system case study illustrates the strengths of the proposed formalism. The requirements meta-model proposed by the CESAR European project is used for requirements engineering while IEC 61131-3 and model-driven concepts are used in the design and development phases. A tool resulting from the proposed framework called REBATE (Requirements Based Automatic Testing Engine) is used to generate and execute test cases for increasingly concrete controller and plant models.*

## 1. INTRODUCTION

Industrial automation software orchestrates multiple mechatronic components through a network of sensors and actuators that control complex physical processes. Examples of these *industrial cyber-physical systems* [1] include airport baggage handling systems, sorting machines, and complex manufacturing systems. Industrial automation systems are safety-critical systems and must be tested thoroughly before being deployed. The life cycle of industrial automation software follows the so-called V-process [2] where design (high-level and then lower-level) and development phases follow requirements engineering as per the traditional waterfall model [3]. The testing phase bends the process steps upwards, as testing happens in a bottom-up fashion, from verification of smaller components up to validation of higher level design and more abstract requirements. Another key aspect of industrial automation software development is the use of *model-driven engineering* (MDE) [4] where physical processes being controlled are abstracted into *a plant model* which is co-developed along with the software. Plant models enable early verification and validation of designs.

Both the V-process model and model-driven engineering emphasize the need for rigorous testing. It is important to test designs as early as possible, as the cost of finding and fixing bugs increases exponentially in later stages. Unfortunately, the levels of details required to test a component, sub-systems, or the whole system are only apparent during or after the development phase. Iterative refinements provide some relief as validation results on early designs can hold on more detailed designs, but requires enormous skills and time from designers and developers [5]. Additionally, time-to-market and cost constraints make maintaining correspondence between designs infeasible.

This paper marries the V-model with MDE, and proposes a novel and elegant formalism to generate test cases for every phase of the design cycle. Fig. 1 shows the proposed modified V-model. An initial requirements *ontology* is developed during the sub-phases of requirements engineering [6]. This ontology specifies, defines, and links system concepts over which requirements in the *system requirements specification* (SRS) are specified. In subsequent phases, such as design, development, and testing, the initial ontology is extended to contain more concrete concepts.

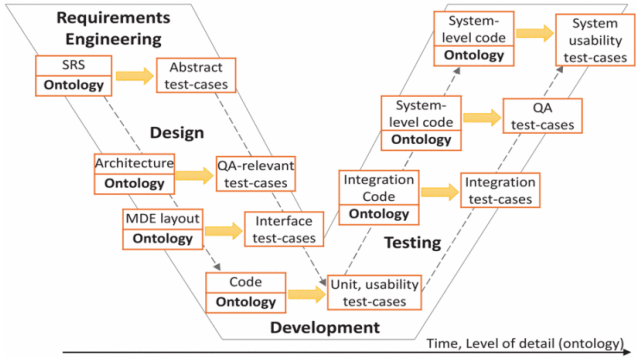

Fig. 1. Modified V-model.

The iterative refinement of ontologies ensures a level of correspondence between various versions of the system. The inter-ontology linkages can be used to automatically generate increasingly concrete test cases for functional requirements related to safety. We propose a simplified relationship between ontologies, requirements, and test cases such that the first two can be used to generate the third. While ontologies must be linked manually, user efforts can be reduced by reusing artefacts created in each stage, and by focusing only on elements such as those related to high-priority safety. Correspondence between safety-relevant aspects is necessary regardless of costs due to certification and assurance demands. The key contributions of this paper are: (a) a general framework to link ontologies and generate test cases for safety related functional requirements, (b) applying this framework to an industry-strength *water process system* (WPS) [7], and (c) the design of the REBATE (*REquirements Based Automatic Testing Engine*) tool for automatically executing test cases.

Our work is similar to the *model-based testing* (MBT) approach [5] for generating test cases. Previously, in [7] the WPS was used to demonstrate the feasibility of MBT in process control applications. In particular, the process control engineering standard IEC 62424 [8] was used to unify various information sources, such as P&IDs and instrument lists, which are then used to build requirement models in UML state charts. A commercial tool then transformed the requirement models into JUnit test cases. REBATE also adopts a similar setup for executing test cases. However, our approach is more generic and does not rely on specific commercial tools. Other works using UML or other formalisms for generating test cases for industrial applications can be found in [9]–[10][11][12]. These works mainly focus on specific stages of the SDLC. In comparison, our work proposes a coherent way of generating test cases for all engineering stages. In [13], an automatic test case generation formalism for web services is presented. Application logic is described using a web ontology language (OWL) variant for web services, and then transformed into a Petri-net model to formalize the structure and behaviour of the service under test. This approach creates more detailed test cases than our current formulation as it also generates test data by reasoning over the ontology. However, it does not provide a progressive refinement of ontology, requirements, and test cases.

## 2. Formalism and Application

The WPS heats, pressurizes, and circulates water to be consumed at a client site. The *piping and instrumentation diagram* (P&ID) of the physical system is illustrated in Fig. 2. We focus on a control loop to maintain the water level of the *Feedwater Tank*. The normal range of the tank level is defined as [**L**, **H**], with an exteme limit of [**LL**, **HH**]. The system must feature appropriate alarms when the normal or extreme ranges are breached and must also display the current tank level. In the initial requirements elicitation stage, we captured two requirements for this control loop:

**R1**. The Feedwater Tank must never overflow or underflow under normal system operations.

**R2**. If the Feedwater Tank overflows or underflows, an alarm must be raised.

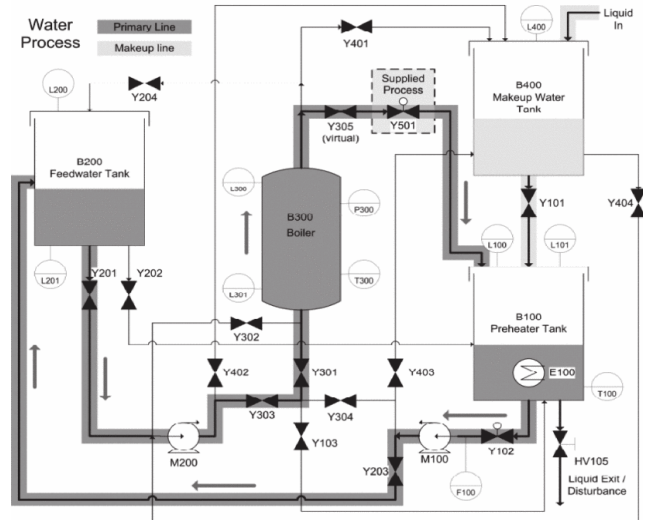

Fig. 2. P&ID diagram of the WPS [7].

We can further break the requirements down into sub-requirements. For example, **R1** can be broken into sub-requirements for maintaining the tank level within [**L**, **H**](**R1_1**), ensuring no overflow (**R1_2**), and ensuring no underflow (**R1_3**). We can use any *requirements specification language* (RSL), for example, guided natural language, boilerplates, or patterns as described in the CESAR project [6]. Boilerplates are semi-complete sentences, such as:

**B1**: When <*state*>**,** then <*system*> never <*state*>**.**

The incomplete parts, like <*system*> and <*state*> called *attributes*, are filled in by requirements engineers. The list of attributes comes from a *requirements ontology*. We define an ontology as a multigraph represented as the tuple $O = <\Sigma_V, \Sigma_A, V, A, s, t, L_V, L_A>$, where $\Sigma_V, \Sigma_A$ are non-intersecting sets of vertex and arc labels respectively, V and A are sets of vertices and arcs

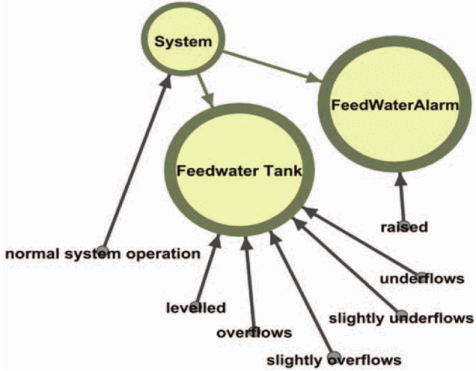

Fig. 3. A simple requirements ontology.

respectively, $s: A \to V$ and $t: A \to V$ are source and target functions mapping each arc to a unique source and a unique target, $L_V: V \to 2^{\Sigma_V}$ and $L_A: V \to 2^{\Sigma_A}$ are the vertex and arc labelling functions respectively. Our definition of ontology is more abstract than OWL as it allows any user-defined relationships between concepts. A minimal ontology needed to specify the requirements captured above is shown in Fig. 3. The larger circles are system concepts. The arrows from **System** to **Feedwater Tank** and **FeedWater Alarm** signify a "contains" relationship. The smaller circles are states of the system components they are linked to. Based on this ontology, we can fill boilerplate B1 to form the following requirement:

**R1_3:** When *System.normal system operation*, then *Feedwater Tank* never *underflows*.

A requirement is defined as a tuple $Req = <O, Q, q_0, L, R, EC, RC>$ where $O$ is an ontology, $Q$ is a finite set of states with $q_0 \in Q$ as a unique initial state, $L: Q \to 2^{\mathcal{B}(\Sigma)}$ labels states with *stay conditions*, $R: Q \times 2^{\mathcal{B}(\Sigma)} \to Q$ is the transition function, $EC \subseteq \mathcal{B}(\Sigma)$ is the set of *entry conditions* and, $RC \subseteq Q \times \mathcal{B}(\Sigma)$ is the set of *release conditions*. Here, $\mathcal{B}(\Sigma)$ refers to the set of all Boolean formulae over $\Sigma$.

A requirement is a finite state machine with a set of *entry conditions* and a set of *release conditions* for each state. The filled boilerplate **R1_3** can be described as a requirement with the only entry condition *System.normal system operation*. The entry condition is shown as an incoming arrow to the only state rs0 of the requirement shown in Fig. 4. The *stay condition* of this state requires the Feedwater tank to not overflow, and is shown as the label of the state. The release condition of the requirement happens when the tank underflows and is shown as an outgoing arrow in the figure.

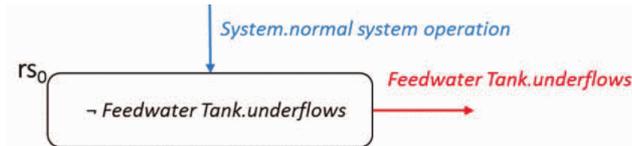

Fig. 4. Safety requirement **R1_3**.

Requirements can be automatically converted into one or more test cases. This is done by traversing the requirement state machine, starting with one of the entry conditions and ending at a release condition. At each step, an entry condition and a staying condition is captured. For requirement **R1_3** shown in Fig. 4, we start with the lone entry condition *System.normal system operation* and pair it with the staying condition ¬*Feedwater Tank.underflows* of the initial state. Hence, (*System.normal system operation,* ¬**Feedwater Tank.underflows**) *represents* the first test-step of a test case. Next, the release condition **Feedwater Tank.underflows** is captured, and the test case ends. The complete test case is:

**R1_3_TC1_V1:**

$(System.normal system operation, \neg Feedwatier\ Tank.undeflows),$

$(FeedwaterTank.underflows.null)$

Formally, given an ontology $O$, a test case is defined as a finite sequence $TC \in TS^* = ts_1, ts_2, \ldots$ where each $ts_i = (PrC_i, PoC_i)$ is a *test-step* where $PrC_i \in \mathcal{B}(\Sigma), PoC_i \in \mathcal{B}(\Sigma)$ are pre- and post-conditions.

At the requirements engineering stage, abstract test cases such as **R1_3_TC1_V1** can be used to find conflicts and infeasible paths between requirements, or to provide clarity about the kind of behaviours a boilerplate requirement would admit. In subsequent stages, more concrete test cases can be generated as the ontology becomes more detailed and requirements can be elaborated further. For example, in the initial high-level design sub-stage of software design of the WPS, we choose a *model-driven architecture*, and apply the model-view-controller (MVC) [14] architectural pattern to divide the systems into plant, controller, and view. We can then further refine any of these parts. For example, the plant can be divided into actuators, sensors, and other components. All these details can be added to the ontology, and then requirements can be updated. We can generate test cases automatically for this more detailed model of the system. The same reasoning can be used at low-level design, development, and testing phases.

## 3. Implementing Automatic Testing (Rebate)

We have implemented the formalism in Java as a library containing formal models of ontology, requirement, test cases, and test case trees. The generic framework was extended and customized for use in the generation and execution of test cases for industrial cyber-physical systems. The resulting tool REBATE uses several third-party tools in addition to the Java library. During the requirements stage, REBATE uses the CESAR tool DODT [15] to formalize requirements and create a requirements ontology, and to ensure completeness, correctness, and consistency of requirements. The requirements ontology and set of requirements can be exported from DODT as OWL and XML files respectively. These files are read into the test case generation library using appropriate parsers. Test cases are then generated automatically using the framework described in the previous section. For design and development phases, the CODESYS tool was used. All artefacts generated during the design and development stages, such as system hierarchy' plant hierarchy, interfaces of individual software components as well as detailed code, were encoded as extensions of the ontology and then imported into the Java library. While this process can be mostly automated, some parts, such as deciding where the additional information generated during design and development stages must be added into the ontology, are guided by the user. During every stage, test cases can be generated and then executed using appropriate tools.

In order to run test cases on the model of the WPS as well as the real plant, we extended an existing tracking simulation environment which runs the controller simultaneously along with the actual plant (process) as well as a simulation plant model [16]. Any differences between their behaviours are recorded and then used to refine the plant model to become more accurate. REBATE automatically transforms test cases generated by our approach into a format (comma-delimited) accepted by the tracking simulation environment.

Fig. 5 shows the results of running three tests relating to the requirements relating to the control loop for maintaining the Feedwater tank level. Test Case 1 checks that the water level in the Feedwater tank (process or model) remained above the **LL** bound even if it went below **L**. As Fig. 5(a) confirms, this test case was met when using both the plant and the simulation model. Test Case 2 checks the case where tank level drops from normal to first between **L** and **LL**, and then to below **LL**. As Fig. 5(b) shows, this never happens in the plant or the model because of extra safety features that deadlock the system when the level is at or below **LL**. Test Case 3 checks that no alarm is raised when the water level remains in the normal range. As Fig. 5(c) shows, while the controller did not raise alarm when the water level reading was set to be very close to **H** by the test case, the actual plant had some oscillations during which the water level went above **H**. This violation of the test case was not visible in the model of the plant.

## 4. Conclusions and Future Directions

We presented a generic and elegant formalism in which all stages of the system development life cycle of industrial cyber-physical systems contribute to the iterative development of a system ontology. Requirements written on an initial ontology created during the requirements engineering phase are made progressively detailed. In our framework, a class of functional requirements that relate to system safety can be automatically converted into test cases. These test cases can be generated at any stage of the system development life cycle and can be used for a variety of purposes, such as testing consistency between requirements during requirements engineering' testing design logic and component interconnections during the design phase and for unit, integration, and system testing during the development and testing stages. A generic tool, called REquirements Based Automatic Testing Engine (REBATE) implements the formalism and is available as open-source software. Future directions of this work include automatically refining ontologies to reduce designer efforts, admitting additional functional and non-functional requirements classes, and automatically managing the impact of changes in requirements and system components.

## ACKNOWLEDGMENT

This work was partially supported by SAUNA and S-STEP projects.

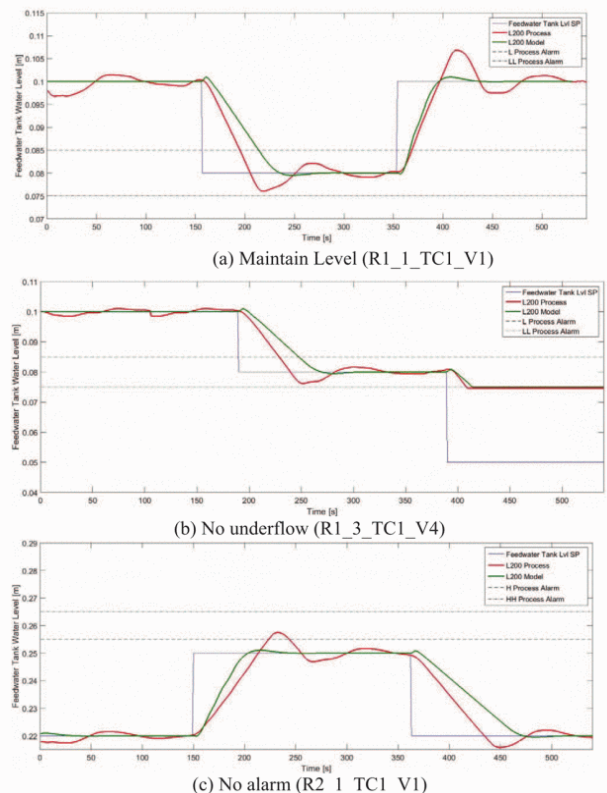

Fig. 5. Results of test case execution.